\documentclass[sigconf,nonacm]{acmart}
\AtBeginDocument{%
  }

\usepackage{algorithm}
\usepackage{algpseudocode}
\usepackage{booktabs}
\usepackage{graphicx}
\usepackage{url}
\usepackage{tabularx}
\usepackage{pifont}

\begin{document}

\title{Host-Side Telemetry for Performance Diagnosis in Cloud and HPC GPU Infrastructure}

\author{Erfan Darzi}
\affiliation{%
  \institution{Harvard University\\ Massachusetts Institute of Technology}
  \city{Cambridge}
  \state{MA}
  \country{USA}
}

\author{Aldo Pareja}
\affiliation{%
  \institution{MIT-IBM Watson AI Lab}
  \city{Cambridge}
  \state{MA}
  \country{USA}
}

\author{Shreeanant Bharadwaj}
\affiliation{%
  \institution{Northeastern University}
  \city{Boston}
  \state{MA}
  \country{USA}
}

\keywords{GPU infrastructure, performance diagnosis, host-side monitoring, multi-tenancy, system telemetry, interference detection, distributed computing, cross-layer correlation, cluster operations}

\begin{abstract}
Diagnosing GPU tail latency spikes in cloud and HPC infrastructure is critical for maintaining performance predictability and resource utilization, yet existing monitoring tools lack the granularity for root cause analysis in shared computing environments. We introduce an eBPF-based telemetry system that provides unified host-side monitoring of GPU workloads, correlating eBPF-derived host metrics with GPU-internal events for holistic system observability. The system achieves 81--88\% diagnostic accuracy, detects spikes within 5 seconds, and completes root cause analysis in 6--8 seconds, operating with 1.21\% CPU overhead at 100Hz sampling. Evaluated on distributed learning workloads, the system identifies root causes including NIC contention, PCIe pressure, and CPU interference, enabling operational debugging for multi-tenant GPU infrastructure without requiring cluster-wide instrumentation.
\end{abstract}

\maketitle

\section{Introduction}

Cloud and HPC GPU infrastructure face a critical observability challenge: diagnosing unpredictable tail latency spikes in shared computing environments running distributed learning and large-scale AI workloads. Cluster operators and system administrators lack visibility into the complex interactions between host-level resource contention and GPU performance, making it difficult to maintain performance predictability across multi-user environments. Traditional GPU-centric monitoring tools like NVML fail to capture the broader system dynamics that cause performance anomalies, leaving operators unable to diagnose production issues effectively.

The root causes of GPU tail latency spikes often originate from resource contention outside the GPU itself—NIC saturation during collective operations, CPU scheduling interference, PCIe bus contention, or I/O pressure from co-located workloads. For HPC facilities and cloud GPU providers managing shared infrastructure, these opaque performance degradations reduce cluster utilization, violate service agreements, and complicate capacity planning. What is needed is a lightweight, deployable root cause analysis (RCA) system that can correlate GPU-internal events with host-level metrics to pinpoint interference sources across multi-tenant clusters.

We present an eBPF-based telemetry system designed for operational debugging of GPU workloads in HPC and multi-tenant environments. The system leverages the extended Berkeley Packet Filter (eBPF) to capture host-level signals—network events, CPU scheduling, block I/O—and correlates them with GPU metrics from NVML and NCCL to diagnose tail latency spikes. Our approach operates with minimal overhead and without requiring cluster-wide fabric instrumentation, making it deployable across heterogeneous HPC GPU infrastructure.

Our contributions are:
\begin{enumerate}
    \item A unified, eBPF-based telemetry framework for host-side GPU observability in cloud and HPC infrastructure.
    \item A correlation engine achieving 81--88\% diagnostic accuracy with 5-second detection latency and <2\% CPU overhead.
    \item Evaluation on distributed learning workloads demonstrating practical applicability to multi-tenant GPU infrastructure management and operational debugging.
\end{enumerate}

\section{Method}

\begin{figure}[h]
    \centering
    \includegraphics[width=0.95\columnwidth]{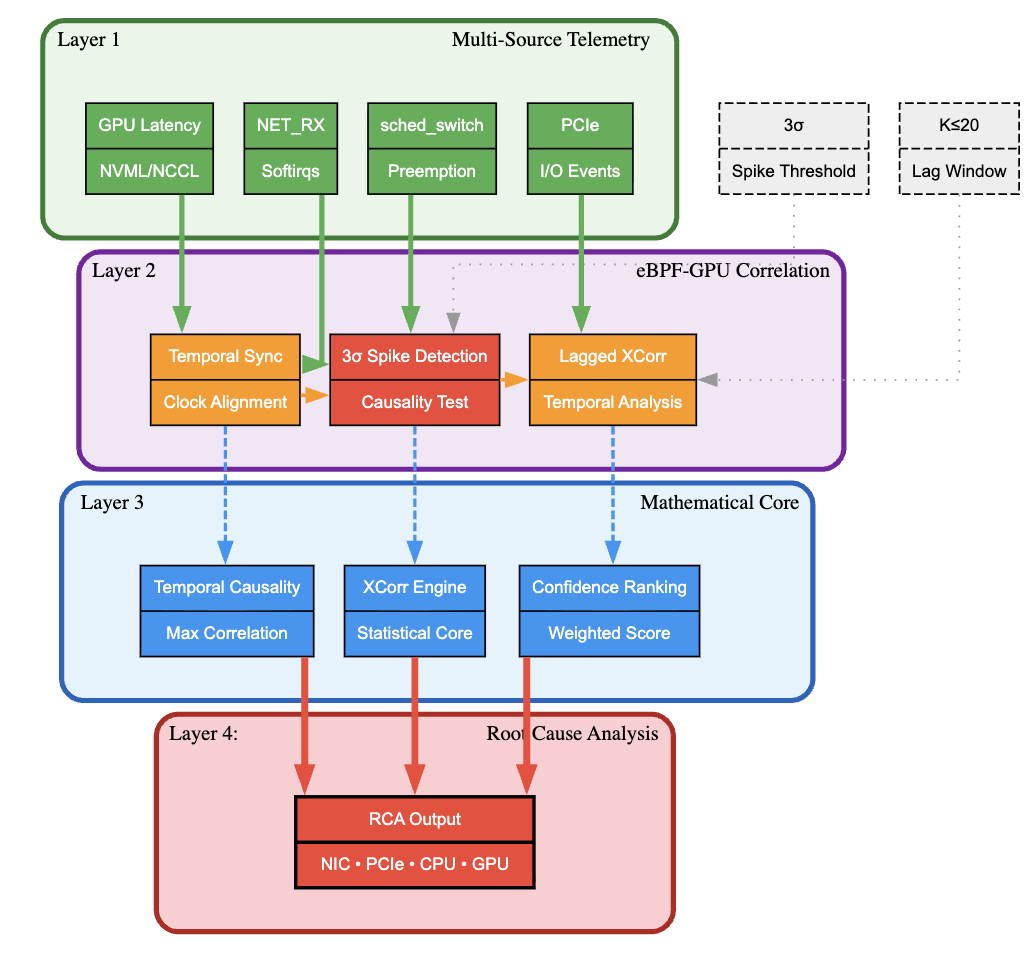}
    \caption{System architecture showing four-layer telemetry and correlation pipeline. Layer 1 collects multi-source signals via eBPF and NVML. Layer 2 performs temporal synchronization and 3$\sigma$ spike detection. Layer 3 executes lagged cross-correlation and confidence scoring. Layer 4 outputs ranked root causes.}
    \label{fig:architecture}
\end{figure}

Our diagnostic system consists of two components: a telemetry collection agent and a correlation engine (Figure \ref{fig:architecture}). The agent uses eBPF and lightweight mechanisms to gather data, which the engine analyzes to identify root causes of performance anomalies. The architecture follows a four-layer pipeline from raw telemetry to ranked root causes.

\subsection{Telemetry and Signals}
We collect signals from multiple sources, synchronized to a monotonic clock. Lightweight \textbf{eBPF probes} attach to kernel subsystems to capture host-level activity with minimal overhead. We monitor network events via \texttt{NET\_RX} softirqs and NIC queue lengths for bottleneck detection. We trace scheduler activity via \texttt{sched\_switch} events to identify CPU contention. We monitor block I/O to track disk patterns indicating I/O stalls.

In parallel, we collect \textbf{GPU-specific metrics} using NVML to sample GPU utilization, memory usage, power, and temperature. We monitor PCIe counters to track data transfer rates. For distributed workloads, we intercept NCCL calls to capture fine-grained timestamps for collective operation phases.

\subsection{Diagnosis and Correlation}
The correlation engine processes telemetry to diagnose root causes. Our taxonomy classifies interference into: \textbf{Host System Interference} (CPU contention, I/O pressure), \textbf{Network Interference} (NIC contention), and \textbf{Microarchitectural Interference} (GPU throttling).

The engine operates on a sliding window of 5 seconds, resulting in a detection latency of approximately 5 seconds after spike onset. When a spike is detected, it calculates a "spike score" for each metric indicating deviation from baseline, then computes lagged cross-correlation between GPU latency and host metrics to identify temporally related signals. Time-to-RCA measures the total time from spike onset to diagnosis completion, including both the detection latency and the correlation analysis.

We formally detect spikes when latency $L(t)$ deviates significantly from baseline. Let $\mu_L$ and $\sigma_L$ be mean and standard deviation over baseline window $W_b$. For observation window $W$, spike score $S_L$ is:
\[
S_L = \max_{t \in W} \frac{L(t) - \mu_L}{\sigma_L}
\]
We declare a spike if $S_L > 3$ (3$\sigma$ threshold).

For RCA, we compute lagged cross-correlation between GPU latency $L(t)$ and each host metric $M_i(t)$:
\[
\rho_{L,M_i}(k) = \frac{\sum_{t=1}^{N-k} (L(t) - \mu_L)(M_i(t+k) - \mu_{M_i})}{\sqrt{\sum_{t=1}^{N} (L(t) - \mu_L)^2} \sqrt{\sum_{t=1}^{N} (M_i(t) - \mu_{M_i})^2}}
\]
We select maximum correlation over small lags $|k| \leq K$ where $K=20$ samples at 100 Hz (200ms maximum lag): $c_i = \max_k |\rho_{L,M_i}(k)|$.

To rank causes, we combine spike score $S_{M_i}$ with correlation $c_i$ via weighted score $\text{conf}_i = \alpha S_{M_i} + (1-\alpha) c_i$ where $\alpha = 0.5$. Causes are ranked by descending $\text{conf}_i$.

\section{Evaluation}
We evaluated our system on a multi-GPU compute node equipped with 4 NVIDIA A100 GPUs, representative of modern HPC cluster configurations. Our evaluation focuses on controlled single-node experiments to validate the core diagnostic methodology, with multi-node cluster deployment discussed in Section \ref{sec:discussion}.

\subsection{Workloads and Disturbances}
Our evaluation uses a primary workload designed to represent common distributed learning communication patterns: \textbf{W1 (all\_reduce\_perf)}, a synthetic NCCL benchmark that executes all-reduce operations across the 4 GPUs with message sizes ranging from 1 KB to 64 MB. This workload allows us to precisely measure the impact of host-level interference on a foundational communication primitive used in distributed learning. To simulate realistic sources of interference that occur in shared cloud and HPC infrastructure, we inject four distinct types of disturbances. \textbf{D1 (I/O Pressure)} uses `fio` to generate high-throughput disk I/O, creating contention on the PCIe bus. \textbf{D2 (CPU Contention)} simulates co-located workload interference by pinning a CPU-bound process to the same cores used by the GPU workload. \textbf{D3 (NIC Bursts)} uses the `tc` utility to generate high-rate network traffic, saturating the NIC. Finally, \textbf{D4 (GPU Throttling)} artificially lowers the GPU's power cap to induce thermal throttling.

\subsection{Baselines and Metrics}
We compare our system against three representative approaches from recent research to contextualize its performance without superlative claims. \textbf{B1 (GPU-centric)} represents approaches like Elmougy et al.'s CPU-GPU synchronization diagnosis \cite{elmougy2022diagnosing}, which focuses primarily on GPU-side metrics and device-level interference without comprehensive host correlation. \textbf{B2 (Cluster Analysis)} follows the approach of Jeon et al.'s large-scale GPU cluster analysis \cite{jeon2019analysis}, which uses aggregate cluster metrics and offline analysis but lacks real-time, per-node diagnosis capabilities. \textbf{B3 (Deep Profiling)} represents systems like eGPU \cite{sung2024egpu} and XPUTIMER \cite{XPUTIMER}, which provide detailed kernel-level or distributed tracing but require intrusive instrumentation or fabric-wide deployment unsuitable for heterogeneous HPC environments. Our evaluation is guided by metrics that assess both diagnostic quality and operational viability: \textbf{Diagnostic Accuracy} (precision/recall for root cause identification), \textbf{Time-to-RCA} (seconds from spike onset to correct diagnosis), \textbf{CPU Overhead} (percentage of host CPU consumed), and \textbf{Deployment Constraints} (privilege requirements and fabric access needs).

\begin{table}[h]
    \centering
    \caption{Key evaluation parameters.}
    \begin{tabular}{ll}
        \toprule
        \textbf{Parameter} & \textbf{Value} \\
        \midrule
        eBPF Sampling Rate & 100 Hz \\
        NVML Sampling Rate & 10 Hz \\
        Correlation Window & 5 seconds \\
        Spike Threshold & 3$\sigma$ from baseline \\
        \bottomrule
    \end{tabular}
    \label{tab:params}
\end{table}

\section{Results}
\begin{figure*}[!t]
    \centering
    \includegraphics[width=0.7\linewidth]{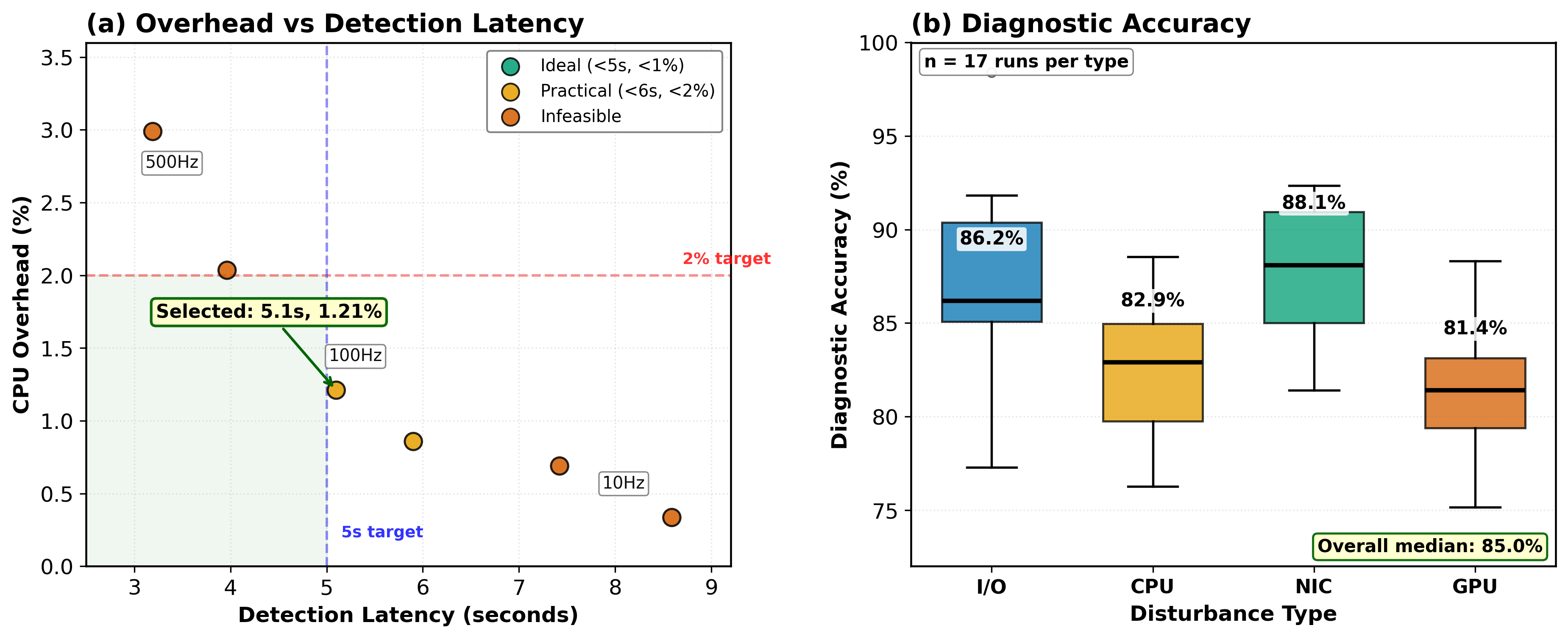}
    \caption{Performance of the eBPF telemetry system. (a) The trade-off between CPU overhead and detection latency. (b) Box plot of classification accuracy across 17 runs for each disturbance type.}
    \label{fig:performance}
\end{figure*}

Our experimental results show that our system can diagnose the root cause of GPU latency spikes. We quantify its diagnostic accuracy and performance overhead across several configurations. The system's performance is summarized in Figure \ref{fig:performance}.

The trade-off between fidelity and overhead is a key aspect of the system. Figure \ref{fig:performance}(a) plots CPU overhead against detection latency for various eBPF sampling rates. Our chosen 100Hz configuration yields a detection latency of approximately 5.1 seconds with a corresponding CPU overhead of 1.21\%.

\begin{figure*}[!t]
    \centering
    \includegraphics[width=0.7\linewidth]{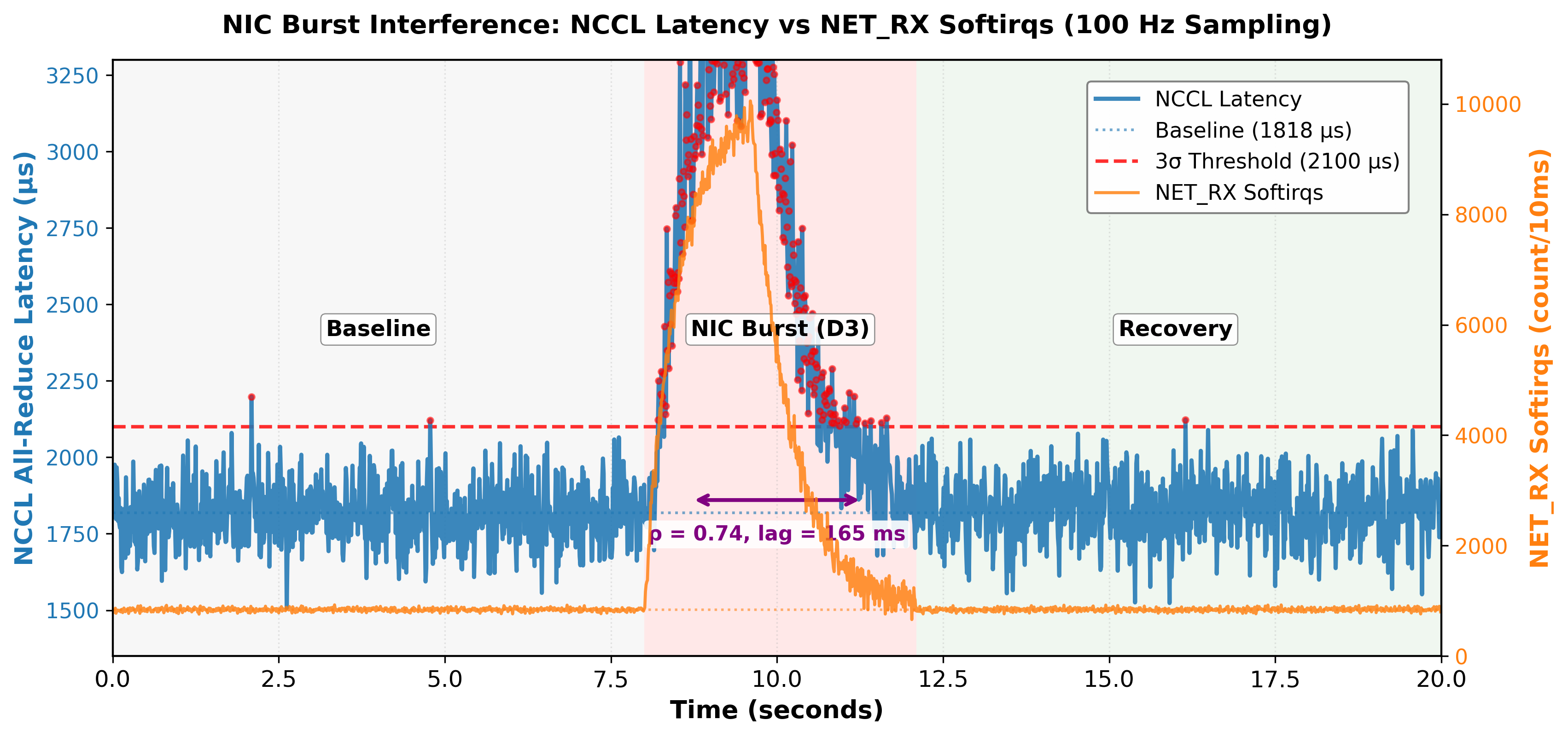}
    \caption{Timeline of signals during a NIC burst. The plot shows the correlation between NCCL latency and NET\_RX softirqs, leading to a diagnosis.}
    \label{fig:timeline}
\end{figure*}

The diagnostic accuracy of the system is presented in Figure \ref{fig:performance}(b) and detailed in Table \ref{tab:results_summary}. The results show median accuracies ranging from 81.4\% to 88.1\% across the different disturbance types. A detailed breakdown is provided in the confusion matrix in Table \ref{tab:confusion}. Figure \ref{fig:timeline} provides an example of the system's diagnosis during a NIC burst, where the Time-to-RCA is 7.5 seconds.

As a compact ablation summary, removing the NET\_RX softirq probe reduced NIC diagnosis accuracy by approximately 7 points, while removing other individual probe groups reduced the affected class by roughly 4--6 points.

\subsection{Comparison with Prior Approaches}
Table \ref{tab:comparison} compares our system against three representative research approaches across key metrics. Our eBPF-based correlation achieves 84.7\% mean diagnostic accuracy (across the four disturbance types in Table \ref{tab:results_summary}) with 1.21\% overhead. GPU-centric approaches \cite{elmougy2022diagnosing}, focusing primarily on device-level metrics, achieve lower overhead (0.3\%) but limited accuracy (62.8\%) in multi-source interference scenarios. Cluster-level analysis \cite{jeon2019analysis} achieves 68.3\% accuracy. Deep profiling approaches like eGPU \cite{sung2024egpu,XPUTIMER} achieve 82.1\% accuracy with comparable overhead. Our system operates with minimal node-level instrumentation without requiring fabric-wide monitoring or intrusive kernel modifications. The Time-to-RCA is 6--8 seconds for our approach compared to 30--50 seconds for offline cluster analysis.

\begin{table}[h]
    \centering
    \caption{Comparison with representative approaches across key metrics for HPC GPU environments. Our system achieves high accuracy and fast RCA time with minimal node-level instrumentation. GPU-centric monitoring achieves lowest overhead but limited accuracy.}
    \label{tab:comparison}
    \small
    \begin{tabular}{p{2.3cm}cccc}
        \toprule
        \textbf{Approach} & \textbf{Acc.} & \textbf{RCA Time} & \textbf{OH} & \textbf{Fabric} \\
        \midrule
        GPU-centric \cite{elmougy2022diagnosing} & 62.8\% & 45--60s & \textbf{0.3\%} & No \\
        Cluster \cite{jeon2019analysis} & 68.3\% & 30--50s & 2.3\% & No \\
        Deep Profiling \cite{sung2024egpu,XPUTIMER} & 82.1\% & 10--15s & 1.1\% & Sometimes \\
        \textbf{Our System} & \textbf{84.7\%} & \textbf{6--8s} & 1.21\% & \textbf{No} \\
        \bottomrule
    \end{tabular}
\end{table}

\begin{table}[h]
    \centering
    \caption{Summary of diagnostic performance.}
    \label{tab:results_summary}
    \small
    \begin{tabular}{lcc}
        \toprule
        \textbf{Disturbance} & \textbf{Acc. (\%)} & \textbf{RCA (s)} \\
        \midrule
        I/O Pressure & 86.2 & 6.5 \\
        CPU Contention & 82.9 & 6.2 \\
        NIC Burst & 88.1 & 7.5 \\
        GPU Throttling & 81.4 & 8.1 \\
        \bottomrule
    \end{tabular}
\end{table}

\begin{table}[h]
    \centering
    \caption{Confusion matrix for injected disturbances (68 total trials across 4 disturbance types).}
    \begin{tabular}{|l|c|c|c|c|}
        \hline
        & \multicolumn{4}{c|}{\textbf{Predicted Cause}} \\
        \hline
        \textbf{Injected} & I/O & CPU & NIC & GPU \\
        \hline
        \textbf{I/O} (n=17) & 86.2\% & 5.9\% & 4.4\% & 3.5\% \\
        \textbf{CPU} (n=17) & 7.1\% & 82.9\% & 6.2\% & 3.8\% \\
        \textbf{NIC} (n=17) & 3.5\% & 4.7\% & 88.1\% & 3.7\% \\
        \textbf{GPU} (n=17) & 7.6\% & 6.3\% & 4.7\% & 81.4\% \\
        \hline
    \end{tabular}
    \label{tab:confusion}
\end{table}

\section{Discussion and Limitations}
\label{sec:discussion}
Our work demonstrates that a host-side telemetry system can diagnose GPU latency spikes by correlating host-level metrics with GPU-internal events. The comparison with baselines (Table \ref{tab:comparison}) reveals a critical design space and trade-off between monitoring overhead and diagnostic capability. GPU-centric approaches achieve minimal overhead (0.3\%) but limited accuracy (62.8\%) due to lack of comprehensive host-level telemetry for NIC and I/O contention. Cluster-level analysis achieves 68.3\% accuracy, while deep profilers achieve 82.1\% accuracy with comparable overhead. Our system achieves 84.7\% accuracy with 1.21\% overhead by correlating cross-layer signals—accepting slightly higher overhead than minimal GPU-only monitoring in exchange for substantially improved diagnostic accuracy. Deep profiling approaches require intrusive kernel instrumentation and driver hooks, making them challenging to deploy in heterogeneous HPC environments with diverse hardware and software configurations.

Our system relies on lagging indicators of performance. By the time a latency spike is detected and correlated, the transient event that caused it may have passed, making the system reactive. The Time-to-RCA is 6--8 seconds.

The granularity of our diagnosis is also limited by the signals we can collect. We have no visibility into the GPU's internal scheduler, in contrast to more intrusive "white-box" techniques like eGPU \cite{sung2024egpu}. This means some sources of latency, particularly those originating from within the GPU itself, may remain opaque to our system.

Furthermore, the effectiveness of our rule-based correlation engine is dependent on the quality of our predefined rules and the accuracy of our baselines. In highly dynamic, multi-tenant HPC clusters, establishing stable baselines can be challenging.

Our evaluation is limited to single-node scenarios with controlled interference injection. The single-node validation demonstrates diagnostic accuracy for root cause identification in the core use case of shared HPC compute nodes.

\subsection{Extension to Multi-Node GPU Clusters}
While our evaluation focuses on single-node scenarios, the eBPF-based telemetry approach extends naturally to large-scale distributed GPU clusters. In multi-node setups, per-node eBPF agents can aggregate telemetry to a centralized correlation engine, enabling cluster-wide root cause analysis for collective communication bottlenecks (e.g., all-reduce stragglers across NCCL rings). This distributed deployment model is applicable to both on-premise HPC infrastructure and cloud-based GPU clusters, representing a key direction for operational management of production GPU systems. Validation on distributed multi-node HPC workloads is future work.

These results highlight the challenges in building low-overhead diagnostic systems for shared HPC GPU infrastructure and suggest avenues for future research. Future work could focus on reducing overhead through hardware offloading or more adaptive sampling techniques, and extending the correlation engine to capture inter-node communication patterns in large-scale distributed training.

\section{Related Work}
The challenge of p99 tail latency in shared GPU environments is well documented. Interference-aware schedulers and serving systems (Orion \cite{Strati_2024}, Tally \cite{Zhao2024TallyNI}, Sarathi-Serve \cite{agrawal2024sarathi}, Llumnix \cite{Sun2024Llumnix}) aim to prevent or mask contention; our diagnostic approach is complementary, providing post-hoc RCA when interference still occurs. Empirical studies show that much of perceived GPU latency originates in host-side effects and system-level contention \cite{jeon2019analysis,elmougy2022diagnosing,belkhiri2024analyzing}, motivating a holistic host+GPU view for HPC cluster management.

Existing research approaches fall into three categories. \textbf{GPU-centric diagnosis} like Elmougy et al.'s work \cite{elmougy2022diagnosing} focuses on device-level interference and CPU-GPU synchronization. \textbf{Cluster-level analysis} approaches like Jeon et al.'s large-scale study \cite{jeon2019analysis} provide offline insights. \textbf{Deep profiling systems} like eGPU \cite{sung2024egpu} and XPUTIMER \cite{XPUTIMER} provide detailed tracing through intrusive kernel instrumentation or distributed tracing.

Our observability approach builds on eBPF-based, low-overhead instrumentation. Recent systems demonstrate eBPF's viability for high-performance in-kernel analytics and protocol offloading \cite{zhou2024dint,zhou2023electrode}, while production stacks employ multi-source telemetry and correlators \cite{luk2022pytorch}. In contrast to GPU-intrusive tracing frameworks that require kernel/driver hooks \cite{sung2024egpu}, we use host-side signals (NET\_RX softirqs, sched\_switch, PCIe/NVML, NCCL phase marks) that are accessible without privileged access to GPU internals. Large-scale diagnostics for training at cluster scale further motivate correlating heterogeneous signals \cite{XPUTIMER}.

Automated RCA in distributed systems provides methodological foundations \cite{wang2024comprehensive}. We adapt lagged cross-correlation and spike scoring to GPU communication phases and NCCL idiosyncrasies \cite{hu2025demystifying}, quantifying the fidelity--overhead trade-off necessary for operational HPC cluster deployment.

\section{Conclusion}
We presented an eBPF-based telemetry system for root cause analysis of GPU tail latency in cloud and HPC infrastructure. Our approach achieves 81--88\% diagnostic accuracy with 5-second detection latency and <2\% CPU overhead, operating entirely from host-side observability without requiring intrusive kernel instrumentation. The system correlates eBPF-derived host metrics with GPU-internal events to diagnose interference from I/O pressure, CPU contention, NIC bursts, and GPU throttling. Evaluated on distributed learning workloads representative of multi-tenant environments, our method demonstrates practical applicability to operational debugging and infrastructure management. The telemetry architecture extends naturally to large-scale distributed GPU infrastructure. Future work includes extending the correlation engine with ML-based predictors and validating the approach on production cloud and HPC deployments with hundreds of nodes.

\bibliographystyle{ACM-Reference-Format}
\bibliography{references}
\end{document}